\documentclass[sigconf]{acmart}
\AtBeginDocument{%
  }
\setcopyright{acmlicensed}
\copyrightyear{2018}
\acmYear{2018}
\acmDOI{XXXXXXX.XXXXXXX}
\acmConference[Conference acronym 'XX]{Make sure to enter the correct
  conference title from your rights confirmation email}{June 03--05,
  2018}{Woodstock, NY}
\acmISBN{978-1-4503-XXXX-X/2018/06}

\usepackage{cleveref}
\usepackage{booktabs}
\usepackage{multirow}
\usepackage{amsmath}
\begin{document}
\title{LEMUR: Large scale End-to-end MUltimodal Recommendation}
\author{Xintian Han}
\authornote{These authors contributed equally to this research.}
\email{hanxintian@bytedance.com}
\affiliation{%
  \institution{ByteDance}
  \city{Hangzhou}
  \country{China}
}

\author{Honggang Chen}
\authornotemark[1]
\email{chenhonggang@bytedance.com}
\affiliation{%
  \institution{ByteDance}
  \city{Beijing}
  \country{China}}

\author{Quan Lin}
\authornotemark[1]
\email{jiangchen.111@bytedance.com}
\affiliation{%
  \institution{ByteDance}
  \city{Beijing}
  \country{China}
}

\author{Jingyue Gao}
\authornotemark[1]
\email{gaojingyue@bytedance.com}
\affiliation{%
 \institution{ByteDance}
 \city{Beijing}
 \country{China}}

\author{Xiangyuan Ren}
\authornotemark[1]
\email{xiangyuanren@bytedance.com}
\affiliation{%
  \institution{ByteDance}
  \city{Beijing}
  \country{China}}

\author{Lifei Zhu}
\email{zhulifei@bytedance.com}
\affiliation{%
  \institution{ByteDance}
  \city{Hangzhou}
  \country{China}}

\author{Zhisheng Ye}
\email{yezhisheng@pku.edu.cn}
\affiliation{%
  \institution{ByteDance}
  \city{Beijing}
  \country{China}}

\author{Shikang Wu}
\email{wushikang@bytedance.com}
\affiliation{%
  \institution{ByteDance}
  \city{Beijing}
  \country{China}}

\author{Xionghang Xie}
\email{xiexionghang@bytedance.com}
\affiliation{%
  \institution{ByteDance}
  \city{Beijing}
  \country{China}}

\author{Xiaochu Gan}
\email{ganxiaochu@bytedance.com}
\affiliation{%
  \institution{ByteDance}
  \city{Beijing}
  \country{China}}

\author{Bingzheng Wei}
\email{bingzhengwei@hotmail.com}
\affiliation{%
  \institution{ByteDance}
  \city{Beijing}
  \country{China}}

\author{Peng Xu}
\email{xupeng@bytedance.com}
\affiliation{%
  \institution{ByteDance}
  \city{San Jose}
  \country{USA}}

\author{Zhe Wang}
\email{zhewang.tim@gmail.com}
\affiliation{%
  \institution{ByteDance}
  \city{Beijing}
  \country{China}}

\author{Yuchao Zheng}
\authornote{Corresponding Author.}
\email{zhengyuchao.yc@bytedance.com}
\affiliation{%
  \institution{ByteDance}
  \city{Hangzhou}
  \country{China}}

\author{Jingjian Lin}
\email{linjingjian000@gmail.com}
\affiliation{%
  \institution{ByteDance}
  \city{Beijing}
  \country{China}}

\author{Di Wu}
\email{di.wu@bytedance.com}
\affiliation{%
  \institution{ByteDance}
  \city{Beijing}
  \country{China}}

\author{Junfeng Ge}
\email{cheng.l@bytedance.com}
\affiliation{%
  \institution{ByteDance}
  \city{Beijing}
  \country{China}}

\renewcommand{\shortauthors}{Han et al.}

\begin{abstract}
Traditional ID-based recommender systems often struggle with cold-start and generalization challenges. Multimodal recommendation systems, which leverage textual and visual data, offer a promising solution to mitigate these issues. However, existing industrial approaches typically adopt a two-stage training paradigm: first pretraining a multimodal model, then applying its frozen representations to train the recommendation model. This decoupled framework suffers from misalignment between multimodal learning and recommendation objectives, as well as an inability to adapt dynamically to new data. To address these limitations, we propose LEMUR, the first large-scale multimodal recommender system trained end-to-end from raw data. By jointly optimizing both the multimodal and recommendation components, LEMUR ensures tighter alignment with downstream objectives while enabling real-time parameter updates. Constructing multimodal sequential representations from user history often entails prohibitively high computational costs. To alleviate this bottleneck, we propose a novel memory bank mechanism that incrementally accumulates historical multimodal representations throughout the training process.
After one month of deployment in Douyin Search, LEMUR has led to a 0.843\% reduction in query change rate decay and a 0.81\% improvement in QAUC. Additionally, LEMUR has shown significant gains across key offline metrics for Douyin Advertisement. Our results validate the superiority of end-to-end multimodal recommendation in real-world industrial scenarios.
\end{abstract}




\maketitle
\section{Introduction}
\begin{figure*}[t]
\centering
  \includegraphics[width=\textwidth]{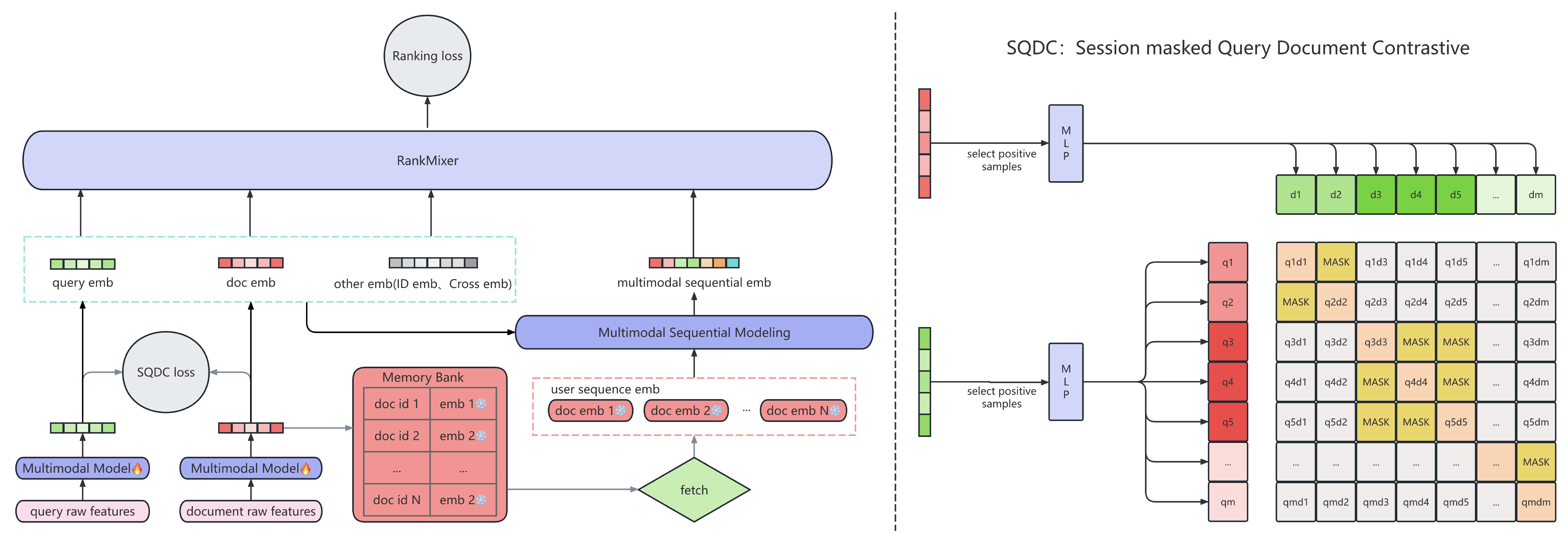}
  \caption{The Overall Framework of LEMUR.}
  \label{fig:structure}
\end{figure*}

In recent years, advancements in hardware—from T4 to A30, and then to A100/H100—coupled with engineering innovations such as Flash Attention~\citep{dao2022flashattention, dao2023flashattention, shah2024flashattention}, Zero Optimization~\citep{rajbhandari2020zero}, and Tensor/Pipeline Parallelism (TP/PP)~\citep{shoeybi2019megatron}, have substantially enhanced the computational capacity of recommendation systems. This progress has enabled the development of more sophisticated network architectures and larger parameter scales, significantly improving the expressive power of models and leading to gains in key performance metrics. Major research and application directions now include scaling up dense neural networks in both parameters and computations~\citep{zhu2025rankmixer, zhang2022dhen, zhang2024wukong, malreddy2024improving, ye2025fuxi, fang2024scaling, zhuang2025practice}, modeling long-range user behavior sequences~\citep{chai2025longer, zhang2024scaling, liu2024kuaiformer, xu2025multi, xu2025efficient, song2025efficient, lyu2025dv365, khrylchenko2025scaling, lai2025exploring}, exploring generative recommendation paradigms~\citep{zhai2024actions, guo2024scaling, liu2024multi, liu2024mmgrec, borisyuk2025features, xiao2025progressive, deng2025onerec, yang2025sparse, huang2025sessionrec, badrinath2025pinrec, zheng2025pre, zheng2025beyond, qiu2025one, zheng2025ega, zhou2025onerec, zhou2025onerec2, kong2025generative}, and advancing multimodal integration~\citep{deng2024end, luo2024qarm, chen2024multi, yan2025mim, li2025bbqrec, liu2020category, chen2016deep, ge2018image, fu2025multimodal, zhou2023bootstrap, zhang2021mining, rajput2023recommender}.

Traditional recommendation models primarily rely on ID-based features, such as user IDs, item IDs, and contextual IDs~\citep{cheng2016wide, rendle2010factorization, wang2021dcn, zhou2018deep}. However, such features inherently suffer from cold-start~\citep{he2016vbpr, mo2015image} and generalization issues~\citep{chen2021efficient}. When a new user joins the system, the absence of historical interactions makes it impossible to generate meaningful recommendations based on ID features alone. Furthermore, since user behavior data is often sparse, models trained solely on ID features tend to exhibit high variance and poor generalization performance.

To address these challenges, multimodal information offers a powerful solution~\citep{deng2024end,luo2024qarm, li2025bbqrec}. By integrating raw text, visual, and other contextual signals, multimodal approaches provide rich and meaningful representations that are not dependent on historical user-item interactions. This is particularly valuable in short-video platforms like Douyin, where content understanding is crucial. Multimodal representations help mitigate the cold-start problem by leveraging content-based features from the very beginning, without waiting for interaction data to accumulate. Moreover, they enhance generalization by capturing semantic information that may not be directly learnable from limited user behavior records.

The current mainstream approach in multimodal recommendation primarily follows a two-stage framework~\citep{mo2015image, yuan2023go, zhou2018deep}. In this setup, an upstream model is first trained on multimodal data to learn multimodal representations (e.g., CLIP~\cite{radford2021learning}, BLIP~\citep{li2022blip}), generating an embedding for each item. These embeddings are then provided as a standalone service to downstream recommendation models. However, a significant gap exists between this two-stage multimodal pre-training and downstream recommendation systems. This misalignment manifests in several ways. First, there is a lack of alignment between content embeddings (derived from multimodal pre-training) and ID embeddings, resulting in content embeddings that are not guided by user behavior information. Second, recommendation models prioritize the learning of ID embeddings, causing most of the informative signals to be stored in ID embeddings and leaving content embeddings underfitted. Third, the update frequencies of multimodal pre-training models and downstream search/recommendation models are inconsistent—while recommendation models are updated in real time, multimodal models often remain fixed after pre-training convergence, leading to a gradual decline in relevance. Finally, when multimodal embeddings are applied to long sequences and transmitted as an online service, severe transmission bottlenecks arise.

In this work, we introduce LEMUR, the first Large-scale, fully End-to-end MUltimodal Recommendation system. Unlike previous approaches such as TIGER~\citep{rajput2023recommender} and EM3~\citep{deng2024end}, which operate under a two-stage training framework and make trade-offs, LEMUR achieves a truly integrated one-stage pipeline that jointly learns multimodal representations and ranking objectives. It stands as the first feasible solution of its kind to be successfully deployed on a large-scale industrial platform. With the memory bank, each training task retains its multimodal sequential representations on a dedicated parameter server, eliminating the need for data transmission from the online service and thereby resolving transmission bottlenecks.

 We posit that as baseline models increase in complexity, the relative resource overhead associated with the end-to-end joint training of recommendation and multimodal models will progressively diminish, making such integrated architectures increasingly practical. LEMUR has been deployed on Douyin Search for more than one month and has achieved significant offline performance improvements in multiple other scenarios. 

Our main contributions are:
\begin{itemize}
    \item We propose the first end-to-end multimodal recommendation system, LEMUR, which has been deployed on the Douyin short video platform. 
    \item We introduce a novel memory bank mechanism to build multimodal sequential representations that saves a huge amount of computations. 
    \item We employ a novel session-masked contrastive loss to align the representation of the query and document. 
    \item We develop sampling techniques to further improve training efficiency.
\end{itemize}

\section{Related Work}
We describe related work in four parts: large recommendation models, multimodal models, multimodal recommendation, and memory mechanisms. 
\subsection{Large Recommendation Models}
Recent advances in hardware capabilities and engineering innovations have ushered recommendation systems into the era of large models. Researchers are expanding the scale of deep neural networks by increasing parameter counts and computational complexity through techniques such as ensemble methods and hierarchical stacking of diverse architectures~\citep{zhang2022dhen, zhang2024wukong, zhuang2025practice}. Zhu et al. introduced a novel architecture named Rankmixer, scaling the total parameters up to one billion~\citep{zhu2025rankmixer}. In parallel, another line of research focuses on extending the length of user behavior sequences to better model long-range dependencies~\citep{chai2025longer, zhang2024scaling, liu2024kuaiformer, xu2025multi, xu2025efficient, song2025efficient, lyu2025dv365, khrylchenko2025scaling, lai2025exploring}. Inspired by the success of next-token prediction in large language models, recent studies have begun to explore next-item prediction in recommendation systems, giving rise to the field of generative recommendation~\citep{zhai2024actions, guo2024scaling, liu2024multi, liu2024mmgrec, borisyuk2025features, xiao2025progressive, deng2025onerec, yang2025sparse, huang2025sessionrec, badrinath2025pinrec, zheng2025pre, zheng2025beyond, qiu2025one, zheng2025ega, zhou2025onerec, zhou2025onerec2, kong2025generative}. Furthermore, multimodal modeling has emerged as a promising direction for scaling up recommendation systems since multimodal encoders typically involve substantial computational overhead. This work integrates the training of such encoders with ranking models, effectively doubling the overall computational load of the recommendation system.

\subsection{Multimodal Models}
The development of base multimodal models paves the way for multimodal recommendation systems. For text, power transformer~\citep{vaswani2017attention} based architectures have achieved a series of SOTA results, including Deberta~\citep{he2020deberta, he2021debertav3}, BERT~\citep{devlin2019bert}, GPT~\citep{radford2018improving} and many others. For vision, convolutional neural networks~\cite{lecun1989backpropagation, krizhevsky2012imagenet, he2016deep} and vision transformers~\cite{dosovitskiy2020image, he2022masked} are the most popular architectures. Many efforts have been made to merge vision and language modalities, including ViLT~\citep{kim2021vilt}, ViLBert~\citep{lu2019vilbert}, CLIP~\cite{radford2021learning}, ALBEF~\cite{li2021align} and BLIP~\cite{li2022blip}. In this work, we follow CLIP and introduce the contrastive loss to align the representation of the query and document. 
\subsection{Multimodal Recommendation}
Multimodal recommendation has received considerable attention in recent research~\citep{deng2024end, luo2024qarm, chen2024multi, yan2025mim, li2025bbqrec, liu2020category, chen2016deep, ge2018image, fu2025multimodal, zhou2023bootstrap, zhang2021mining, rajput2023recommender, mo2015image}. Many existing approaches adopt a two-stage pipeline, where the multimodal encoder is pre-trained offline and its frozen features are subsequently fed into the recommendation model~\cite{yan2025mim, liu2020category, mo2015image, fu2025multimodal, zhou2023bootstrap, zhang2021mining, chen2024multi}.

To reduce computational overhead, some studies opt to train only a subset of the multimodal encoder parameters. For instance, Deng et al.~\cite{deng2024end} fine-tune a q-former module while keeping the vision and language encoders fixed. Similarly, Ge et al.~\cite{ge2018image} update only the top layers of the encoder, leaving the remaining layers frozen. Another line of work converts multimodal content into discrete token IDs through quantization or tokenization techniques~\cite{luo2024qarm, li2025bbqrec, rajput2023recommender}. However, such discrete representations may suffer from information loss relative to the original images or texts. In contrast, our approach trains the multimodal encoder end-to-end alongside the ranking model.

Deng et al. also introduced a Content-ID Contrastive (CIC) learning loss to align frozen multimodal representations with ID embeddings, aiming to produce more task-aware multimodal features and more generalized ID embeddings~\cite{deng2024end}. Nevertheless, our experiments indicate that when the multimodal encoder is updated jointly with the ranking model, the CIC loss does not yield additional performance gains.

Chen et al.~\cite{chen2016deep} explored training a CNN jointly with the ranking model on a real-world dataset. To reduce computational cost, they grouped samples sharing the same image within the same batch. However, this strategy has not been widely deployed and is incompatible with online training pipelines. Moreover, their method does not adequately address multimodal sequence modeling. In this work, we propose a novel cross-worker deduplication technique and an efficient sampling strategy to accelerate training, enabling full deployment in Douyin Search. We further introduce a memory bank mechanism to effectively construct multimodal sequence representations.

\subsection{Memory Mechanisms}
Many researchers leverage memory mechanisms to store and retrieve information for subsequent use. For instance, He et al.~\cite{he2020momentum} introduce a memory bank to store past key representations, which are later sampled to facilitate momentum contrastive learning. Similarly, Key-Value Memory Networks~\cite{miller2016key} employ a memory structure to retain key-value pairs as a knowledge base, thereby improving machine comprehension of documents. Lample et al. further enhance this approach by decomposing keys and values for greater efficiency. More recently, Lu et al.~\cite{lu2025large} incorporate memory network mechanisms to advance recommendation systems.
\section{Methods}
The overall architecture of LEMUR is illustrated in \cref{fig:structure}. Firstly, LEMUR encodes multimodal information using a bidirectional transformer. In contrast to previous works that rely on frozen representations, our transformer is jointly trained with the ranking model. To learn a better multimodal representation, we introduce a noval contrastive learning framework SQDC (Session masked Query to Document Contrastive). In each batch, the query and document representations use real-click signals to filter positive samples for contrastive learning. Furthermore, we employ a memory bank mechanism to store multimodal representations and construct a multimodal sequential representation, thereby reducing computational costs. The multimodal representations together with other features are sent to RankMixer~\citep{zhu2025rankmixer} to produce the final logit. We provide a detailed description of the overall framework below.
\subsection{Problem Formulation}
Our goal is to predict the click-through rate (CTR) in Douyin Search based on features from the query, the user, and the video (also referred to as the document (doc)). In this context, CTR is defined as the probability that a user watches a recommended video for more than five seconds after submitting a query. The model is optimized using the binary cross-entropy loss. Denote $x$ as the input and $y$ as the label from the dataset $\mathcal{D}$. The estimated CTR is $\hat{y} = f_{\theta}(x)$. The binary cross-entropy loss is defined as 
\begin{equation}
    \ell_{\text{CTR}} = -\frac{1}{|\mathcal{D}|}\sum_{(x,y) \in \mathcal{D}}[y\log \hat{y} + (1-y) \log(1-\hat{y})]
\end{equation}
Although additional engagement signals such as likes and comments, are also considered in practice, these are not the primary focus of this work and are omitted for clarity.
\subsection{Raw Features Modeling}
For document i, we employ a pair of bidirectional transformers~\cite{vaswani2017attention} to encode the query raw features $q_{\text{raw}}^{i}$ and document raw features $d_{\text{raw}}^{i}$ separately.
\begin{align}
    q_{i} &= \text{Transformer}(q_{\text{raw}}^{i}) \\
    d_{i} &= \text{Transformer}(d_{\text{raw}}^{i})
\end{align}
A distinct $\texttt{cls}$ token is inserted at the start of both the query and the document sequences. Within the document input, $\texttt{sep}$ tokens are placed between each feature to demarcate their boundaries. In addition to standard position embeddings, we incorporate type embeddings to distinguish among different feature categories. The final representations for query and document are derived from the output corresponding to its $\texttt{cls}$ token. These query and document representations are then concatenated with other features and passed through RankMixer to generate the output logit.
\subsection{SQDC}
To align the representation of the query and the doc, we use an extra in-batch session-masked query-document contrastive (SQDC) loss. 

In a batch of size $K$, denote $q_i$ as the embedding of the $i$-th query and $d_i$ as the embedding of the $i$-th doc. For a positive sample $i$, the contrastive loss aims to bring $q_i$ closer to $d_i$ than any other $d_j, \forall j\neq i$. We use a cosine similarity score to measure the distance between the query and the doc embedding. The cosine similarity is defined as 
\begin{equation}
    \text{sim}(q,d) = \frac{q^T d}{\|q\|\|d\|}
\end{equation}
And the contrastive loss is defined as follows:
\begin{align} 
    \ell_{\text{contrastive}}&=\sum_{i, \text{label}_i = 1} -\log \frac{\exp\left(\text{sim}(q_i,d_i) * T\right)}{\sum_{j=0}^K{\exp\left(\text{sim}(q_i, d_j) * T\right)}} 
\end{align}
 where $T$ is a temperature hyperparameter. 
 
 In the same batch, several samples may come from the same session, belonging to the same query. Therefore, when calculating the contrastive loss, we introduce session-level sample masks to mask samples from the same query. Denote the mask as $A_{ij}$ and QID as the query identifier. The SQDC loss is defined as 
 \begin{align} 
    \ell_{\text{SQDC}}&=\sum_{i, \text{label}_i = 1} -\log \frac{\exp\left(\text{sim}(q_i,d_i) * T\right)}{\sum_{j=0}^K{A_{ij} * \exp\left(\text{sim}(q_i, d_j) * T\right)}} \\ 
    A_{ij} &= \begin{cases} 
    0 & \text{if } i \neq j \text{ and } \text{QID}_i=\text{QID}_j \\
    1 & \text{else} \\
    \end{cases}
\end{align}
We illustrate SQDC in the right panel of \cref{fig:structure}. The yellow colored masks are the session-level masks, i.e., $q_1, q_2$ belong to the same query and $q_3, q_4, q_5$ belong to the same query. The session-level mask stabilizes the training, as the user behaviors from the same query are highly correlated.
\subsection{Memory Bank}
In addition to generating multimodal representations for the target document, we aim to construct multimodal sequential representations for the user’s historical documents. However, jointly training these sequential representations with the recommendation model is computationally prohibitive. For instance, with a batch size of 2048, the base model consumes 1.6 TFLOPs, while the document transformer requires 2.3 TFLOPs. If we consider a sequence length of 1024 or more, the computational cost would increase a thousandfold.

To mitigate this expense, we employ a memory bank strategy (see the memory bank module in~\cref{fig:structure}). During each training batch, we first store the target document’s representation in the memory bank. Then, for any document in the user’s history sequence, we retrieve its corresponding multimodal representation from the memory bank using the document ID. This approach allows us to efficiently form the sequential representation of historical documents. Since we only consider user history within a one-month window and our training dataset spans over two months, the memory bank is sufficient to cover all document representations encountered in the sequence.

\subsection{Multimodal Sequential Modeling}
We employ a modified version of LONGER~\cite{chai2025longer}, a Decoder module, and a similarity module to model the multimodal sequence derived from the memory bank.

In the first layer, the Decoder first transforms other features into a Decoder query token \( Q_1 \) and applies a cross-attention mechanism. Let the multimodal sequential representations be denoted as \( d_1, d_2, \dots, d_N \). These representations serve as keys and values in the cross-attention operation. The output of the cross-attention is then passed through a feed-forward network (FFN) to generate the query token for the next layer. The overall process is summarized below and illustrated in \cref{fig:decoder}:
\begin{figure}
    \centering
    \includegraphics[width=\linewidth]{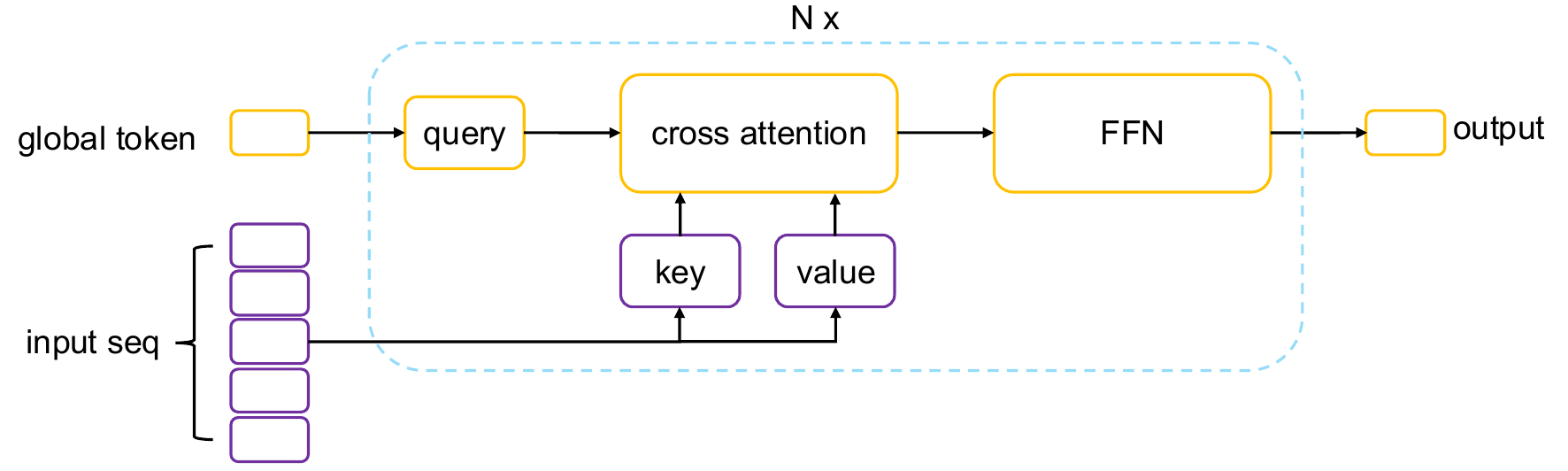}
    \caption{The Structure of Decoder}
    \label{fig:decoder}
\end{figure}

\begin{align}
&Q_{i+1} = \text{FFN}\left(\text{Cross Attention}(Q_{i}, d_1, d_2, \dots, d_N)\right), \\
&\text{Cross Attention}(q_{i}, d_1, d_2, \dots, d_N) = \sum_{j=1}^N a(Q_{i}, d_j) d_j, \\
&a(Q_{i}, d_j) = \frac{\exp(Q_i^T d_j)} {\sum_{j=1}^N \exp(Q_i^T d_j)}.
\end{align}

The Decoder applies only the FFN to the cross-attention output, which significantly reduces computational overhead. This Decoder architecture is also used as the sequence modeling component in the baseline model.

Additionally, a similarity module computes cosine similarities between the target document and each document in the history. To stabilize training, we also maintain a ranked version of the resulting similarity vector.

The outputs from the Decoder and the similarity module are combined with other features and serve as input to the RankMixer model. We have multiple sequences in the model. The longest one has 1000 items. 

\subsection{Efficiency Optimization}
Although we used the memory bank to save a large part of the computational burden, we still have more FLOPS in the Transformer than the ranking model itself. To further speed up training and serving, we apply flash attention~\citep{dao2022flashattention} and mixed-precision training. 

Additionally, we randomly select a subset of training samples to execute the forward and backward passes of the transformer. For the remaining samples, we reuse the embeddings stored in the memory bank instead of performing the forward pass. To balance computational cost and performance, we run both forward and backward processes on 20\% of the samples. During inference, embeddings are directly retrieved from the memory bank without activating the transformer.

Furthermore, a single batch often contains multiple duplicate documents. Since the representations of identical documents can be shared within the batch, we apply a deduplication technique to retain only one unique copy of each document per batch across all the workers before processing with the Transformer. This approach significantly reduces computational overhead. See \cref{sec:dedup} for details. 
\section{Experiments}

\subsection{Experiment Settings}
\subsubsection{Datasets}
To evaluate the performance of our method, we conduct offline experiments using data sourced from Douyin's search recommendation system. The dataset is derived from a subset of online user interaction logs spanning 70 consecutive days and comprises 3 billion samples. Training batches are organized around multiple identical queries, where each query and the corresponding retrieved documents appear consecutively in the training samples. Each sample incorporates user features, query features, document features, cross-features, and user search history features, encompassing billions of user IDs and hundreds of millions of document IDs. Beyond traditional features, we also integrate raw textual information from both queries and documents. The document text includes Optical Character Recognition (OCR) results, Automatic Speech Recognition (ASR) transcripts of the video content, video titles, and OCR text extracted from video cover images. An internal tokenizer is employed to convert raw text into tokens. A summary of the text inputs is provided in \cref{tab:input}.
\begin{table}[h]
    \centering
    \caption{Summary of the text input}
    \begin{tabular}{c|ccc}
    \toprule 
      \textbf{source}   & \textbf{feature} & \textbf{max length}& \textbf{avg length} \\ \midrule
     query    & query  &  10 & 4 \\ \midrule
     \multirow{4}{*}{doc} &title  &31 &22 \\
     &OCR  & 124& 81 \\
     &ASR & 12 & 2 \\
     &cover image OCR & 12 & 6 \\ \bottomrule
    \end{tabular}
    \label{tab:input}
\end{table}
\subsubsection{Baselines}
We conducted experiments on two state-of-the-art production models in our system:
\begin{itemize}
    \item RankMixer~\cite{zhu2025rankmixer}: A high MFU recommendation model with token mixing and per-token FFN  strategies to capture heterogeneous feature interactions efficiently.
    \item LONGER~\cite{chai2025longer}: A long-sequence optimized transformer structure for GPU-efficient recommenders.
\end{itemize}

\subsubsection{Evaluation Metrics}
For offline evaluation, we employ AUC (Area Under the Curve) and QAUC (Query-level AUC). QAUC calculates the AUC for samples within each query and then averages the results across all queries. We also report the number of dense parameters and training FLOPs, with the latter computed using a batch size of 2048.

For online evaluation, we use the query change rate, which measures the probability of users manually refining a search query into a more specific one—for example, changing from “tank” to “tank 300”. It is calculated as follows:
\begin{align}
    &\text{query change rate} =  \nonumber \\
    & \frac{\text{number of distinct UID-query pairs with query reformulation}}{\text{total number of distinct UID-query pairs}}.
\end{align} This metric serves as an indicator of a negative search experience for users.

\subsubsection{Implementation Details}
For the convenience of our experiment, the recommendation model is loaded with checkpoint, and the multimodal model is cold started. All baselines utilize the same batch size, and optimizer settings. The batch size is set to 2048. The dimension of query and document is set to 128 in the memory bank. Since QID is not repeated, the query Transformer layer is set to 2 for serving efficiency. The document Transformer layer is set to 4 for better performance.

\subsection{Offline Evaluation}
We present a comparison of LEMUR with our online baseline model, RankMixer+LONGER. To further demonstrate the effectiveness of LEMUR, we also include the improvements achieved by RankMixer and LONGER individually over an older baseline, DLRM-MLP~\cite{naumov2019deep}. The results are summarized in \cref{tab:model_performance}. Our proposed method LEMUR-SQDC achieves a QAUC improvement of 0.47\%, while LEMUR-SQDC-MB improves QAUC by 0.81\%. In comparison, RankMixer alone yields a 0.59\% gain in QAUC, and LONGER alone contributes a 0.45\% improvement. It is worth noting that in our industrial setting, an improvement of just 0.1\% is considered significant enough to impact online A/B test performance. Therefore, LEMUR delivers a substantial improvement in QAUC.

\begin{table}[htbp]
  \centering
  \caption{Model performance compare with online baselines}
  \begin{tabular}{l c c}
    \toprule
 \textbf{model}  & \textbf{AUC}$\uparrow$ & \textbf{QAUC}$\uparrow$ \\
    \midrule
    DLRM-MLP & 0.74071 & 0.63760 \\
    \midrule
    \quad+RankMixer & +0.49\% & +0.59\%  \\
    \quad\quad+LONGER & +0.89\% & +1.04\%  \\
    \midrule
    \quad\quad\quad\textbf{+LEMUR-SQDC} & +1.22\% & +1.51\% \\
    \quad\quad\quad\textbf{+LEMUR-SQDC-MB} & +1.44\% & +1.85\% \\
    \midrule
    Improvement & 0.55\% & 0.81\%  \\
    \bottomrule
  \end{tabular}
  \label{tab:model_performance}
\end{table}

\subsection{Online A/B Test}
In a recent 14-day A/B test conducted on Douyin Search, LEMUR demonstrated a query change rate of 0.843\%, outperforming the baseline model. The results were statistically significant according to a T-test. Following this successful experiment, LEMUR has been fully deployed on Douyin Search and has been operating effectively for over one month.

\subsection{Comparison with the Two-Stage Method}
We compared LEMUR with a two-stage approach that involves pretraining the transformer with the same SQDC loss and text features on a one-month data sample from Douyin Search, and then using the frozen representations. In the downstream recommendation task, the multimodal representations generated by the pretrained transformer were utilized in the same manner as in LEMUR. As shown in \cref{tab:method_auc}, LEMUR outperforms the two-stage method by 0.69\% in QAUC.

\begin{table}[!h]
  \centering
  \caption{Comparison with the Two-Stage Method}
  \begin{tabular}{l c c}
    \toprule
    \textbf{method} &  \textbf{QAUC} & \textbf{$\Delta$QAUC} \\
    \midrule
    baseline   & 0.64393  & -- \\
    Two-Stage   &  0.64470  & 0.12\% \\
    LEMUR   & 0.64914 & 0.81\% \\
    \bottomrule
  \end{tabular}
  \label{tab:method_auc}
\end{table}

\subsection{Staleness and Coverage}
\begin{figure}[!h]  
  \centering 
  \includegraphics[width=\linewidth]{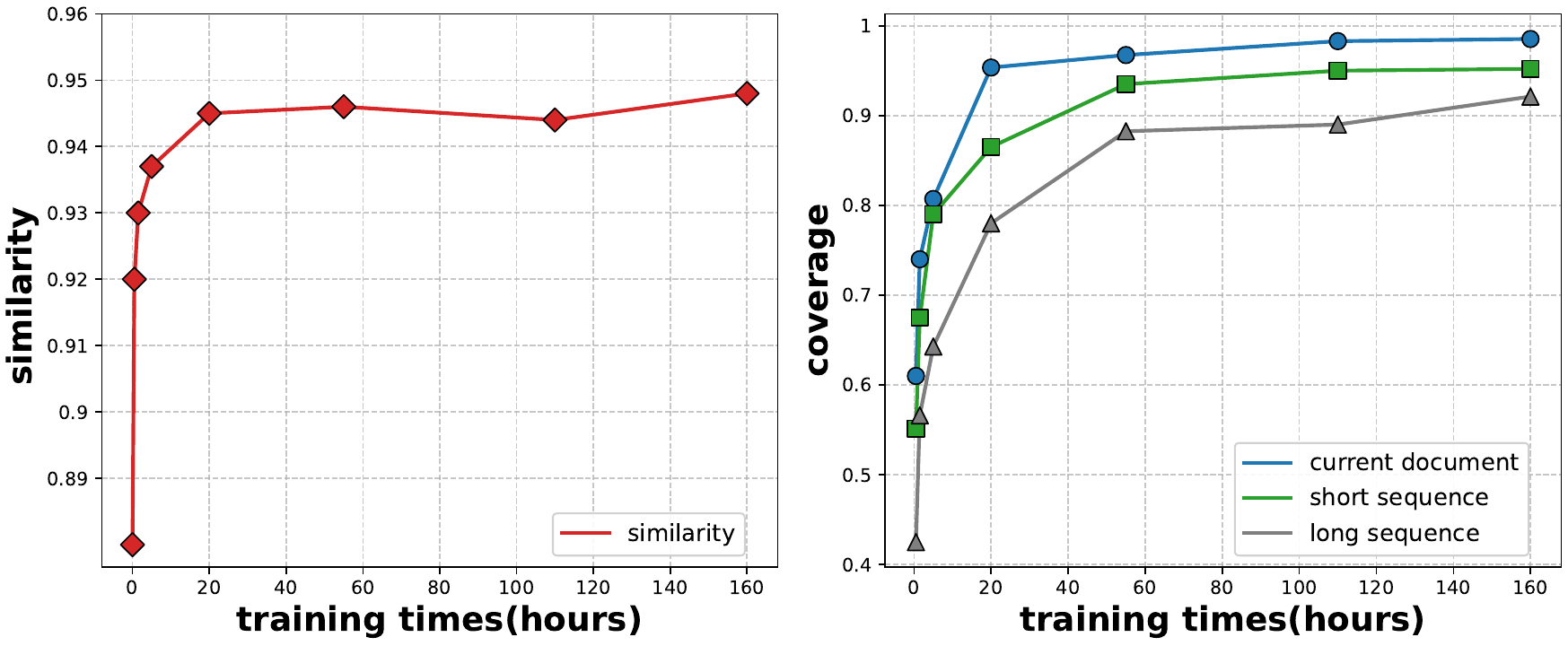} 
  \caption{Staleness and coverage in memory bank.} 
  \label{fig:Staleness} 
\end{figure}

The current memory bank mechanism faces two primary challenges: staleness and coverage. Staleness arises from the discrepancy between the cached representations in the memory bank and the outputs of the current model, as the model parameters are updated during training while the stored representations are not refreshed in real time. We measure staleness using the similarity between the current model’s output and its cached version—treating higher similarity as an indicator of lower staleness. As shown in the left panel of \cref{fig:Staleness}, this similarity increases steadily during training and eventually stabilizes around a high value of 0.95, indicating that staleness does not pose a significant issue for LEMUR.

Coverage, on the other hand, refers to the percentage of sequence items stored in the memory bank relative to the total number of items in the sequence. High coverage is essential for effective sequence modeling. The right panel of \cref{fig:Staleness} illustrates that the short sequence achieves high coverage early in training, while the long sequence requires more time to reach a comparable level. In both cases, coverage exceeds 90\%. Additionally, we visualize the coverage rate for the currently stored document in the memory bank, which remains above 98\%—confirming that embeddings retrieved from the memory bank are suitable for serving.

\begin{table}[!h]
  \centering
  \caption{Performance \& efficiency comparison of different sample training strategies}
  \begin{tabular}{l c c c}
    \toprule
    \textbf{method} & \textbf{params}  &  \textbf{FLOPs} & \textbf{QAUC}  \\
    \midrule
    baseline       & 92M & 1673G & -- \\
    LEMUR, no-sampling  & 139M & 4303G & +0.81\%\\
    LEMUR, p=100,q=20  & 139M & 3232G & +0.78\%\\
    LEMUR, p=20,q=20  & 139M & 2439G & +0.76\%\\
    LEMUR, p=10,q=10  & 139M & 2262G & +0.75\%\\
    \bottomrule
  \end{tabular}
  \label{tab:sample_auc}
\end{table}

\begin{table*}[!h]
    \centering
    \caption{Ablation Studies on LEMUR}
    \begin{tabular}{c|ccc}
    \toprule 
     \textbf{method}    & \textbf{setting} & \textbf{QAUC} & \textbf{$\Delta$QAUC} 
     \\ \midrule
     LEMUR   &  final online version &  +0.81\%  & --
     \\ \midrule
     Transformer & w/ stop gradient in RankMixer & +0.50\%  & -0.31\%
     \\ \midrule
     \multirow{3}{*}{MemoryBank}    & w/o short multimodal sequence  &  +0.77\%  & -0.09\% \\ & w/o long multimodal sequence  &  +0.56\%  & -0.25\%  
     \\ & w/o cosine similarity  &  +0.69\%  & -0.12\% \\\midrule
     \multirow{3}{*}{Loss} 
                 & w/o SQDC Loss & +0.64\% & -0.17\% \\
            & w/o session-level mask & +0.71\% & -0.10\% \\ 
     & w/ CIC Loss & +0.79\% & -0.02\% \\ 
     \bottomrule
    \end{tabular}
    \label{tab:lemur_ablation}
\end{table*}

\begin{table*}[!h]
  \centering
  \caption{Different temperatures and negative samples strategies of SQDC.}
  \begin{tabular}{l c c }
    \toprule
    \textbf{configuration}  & \textbf{QAUC} & \textbf{$\Delta$QAUC}\\
    \midrule
    SQDC (different temperature) & &   \\
    \quad temperature=1 & 0.64831  & +0.68\% \\
    \quad temperature=5  & 0.64857 & +0.72\% \\
    \quad temperature=50  & 0.64914 & +0.81\% \\
    \quad temperature=100 & 0.64869 & +0.74\% \\
    \midrule
    SQDC (different negative samples strategy) &  &  \\
    \quad in-batch positive samples & 0.64914 & +0.81\% \\
    \quad in-batch samples & 0.64908 & +0.80\% \\
    \bottomrule
  \end{tabular}
  \label{tab:config_performance}
\end{table*}

\subsection{Ablation Studies}
We systematically ablated the QAUC gains from different components. 

\subsubsection{The Effect of Joint Training} In RankMixer, we apply gradient stopping to the representations learned from the Transformer. As shown in \cref{tab:lemur_ablation}, this results in a 0.31\% decrease in QAUC, indicating that the relevant features and signals from recommendation models can effectively guide the training of multimodal models and enhance CTR prediction.

\subsubsection{The Usage of Memory Bank} As shown in \cref{tab:lemur_ablation}, the use of multimodal long sequences and short sequences leads to QAUC improvements of 0.09\% and 0.25\%, respectively, underscoring the potential of integrating ultralong multimodal sequences into recommendation systems. Additionally, adopting cosine similarity helps the model better capture content similarity between the current document and historical documents, which contributes to a further 0.12\% gain in QAUC.

\subsubsection{Different Sampling Strategies} We evaluate various training strategies while taking computational cost into account, as shown in \cref{tab:sample_auc}. The FLOPs of the unoptimized LEMUR model are more than doubled, making such a sharp rise in training cost impractical. To mitigate this, we only perform forward and backward passes on a subset of the training data. Let $p$ denote the percentage of samples used in the forward pass and $q$ the percentage used in the backward pass. 

Initially, we conducted forward propagation on all samples in each batch while performing backward propagation on only 20\% of them ($p$ = 100, $q$ = 20 ). This led to a slight drop in QAUC of 0.03\%, suggesting that even after training stabilizes, backward propagation remains important in LEMUR. 

We then attempted both forward and backward propagation on 20\% of the samples in each batch, retrieving the remaining 80\% from the Memory Bank (\( p = 20, q = 20 \)). This resulted in a QAUC decrease of 0.05\%, indicating that data staleness also affects model performance. 

Finally, we further reduced the sample rate to 10\% for both forward and backward passes, which led to an additional 0.01\% decline in QAUC. Given that the 20\% sampling strategy already meets the ROI requirement and offers better coverage of multimodal long sequences, we adopted it for final deployment.

\subsubsection{Analysis of SQDC}
We begin by ablating the use of the SQDC loss and session-level masking. As shown in \cref{tab:lemur_ablation}, the SQDC loss contributes a 0.17\% improvement in QAUC, with the session-level mask accounting for 0.10\% of that gain.

We further examine the effect of the temperature hyperparameter. Consistent with findings in multimodal pre-trained models~\cite{he2020momentum, wang2021understanding}, a higher temperature sharpens the distinction between similarity scores, allowing the model to focus more effectively on positive pairs—such as matched query-document pairs—and thereby improving the aggregation of semantically similar samples. After testing a range of values, we identified 50 as the optimal temperature, which yields a 0.81\% increase in AUC. Both excessively high and low temperatures are observed to degrade performance.

In addition, we explore negative sampling strategies for contrastive learning. As presented in \cref{tab:config_performance}, using either all other in-batch samples or all other positive samples as negatives leads to only a marginal difference in QAUC.

\subsubsection{CIC Loss}
We also explore the use of the CIC loss~\cite{deng2024end}. As shown in \cref{tab:lemur_ablation}, introducing the CIC loss leads to a slight performance degradation under LEMUR's joint training framework. This suggests that explicit alignment with ID embeddings may offer limited benefits when multimodal embeddings are co-optimized with the ranking model.

\subsection{Performances in Other Scenarios}
LEMUR has also achieved favorable offline and online gains in other scenarios. Specifically, on the Douyin Ads Platform, we obtain 0.1\% offline AUC gain. In the TikTok Search scenario, the online A/B test gain approximately achieves a 0.843\% reduction in query change rate.

\section{Discussion}
In this work, we introduced LEMUR, the first fully end-to-end multimodal recommendation system successfully deployed at a large industrial scale. By integrating the learning of multimodal representations and ranking objectives into a single, joint training pipeline, LEMUR overcomes the limitations and trade-offs inherent in previous two-stage frameworks. Our key innovations—including a novel memory bank mechanism for efficient sequential modeling, a session-masked contrastive loss for query-document alignment, and advanced deduplication and sampling techniques—collectively enable this end-to-end paradigm with high training efficiency. The successful deployment of LEMUR on Douyin Search for over one month, along with significant offline gains in other scenarios, validates its effectiveness and practical viability. We believe that as model complexity continues to grow, the integrated architecture exemplified by LEMUR represents a promising and increasingly practical direction for the future of multimodal recommendation.

Despite the demonstrated success of LEMUR, we acknowledge certain limitations in our current design, which open avenues for future research. A primary point of discussion revolves around the memory bank mechanism. While it is instrumental in saving computational costs for building sequential recommendations, it introduces an inherent issue of staleness. The representations stored in the memory bank are not updated in real-time as the model parameters evolve during training, leading to a potential discrepancy between the cached representations and the current model's output. This staleness could hinder the optimization process and the quality of the learned sequential patterns. Furthermore, the coverage and update strategy of the memory bank may not fully capture the long-tail distribution of items in a large-scale system, potentially biasing the model towards popular items. Addressing these challenges related to staleness and coverage will be a critical focus of our future work, potentially through exploring dynamic update schedules or more sophisticated caching architectures.

\begin{acks}
We thank Gangqin Zhuang, Kaiyuan Ma, Zunyao Xue, Tianyi Liu, Minghui Yang, Hao Jiang, Tianlong Fan, and Wenda Liu for their help on efficiency optimization. 
\end{acks}

\clearpage
\bibliographystyle{ACM-Reference-Format}

\appendix
\section{Deduplication\label{sec:dedup}}
We reuse the computation of document embeddings. Within a single training step, embeddings computed on one rank are made available to all other ranks for immediate local reuse. This is achieved using the \textit{all\_gather} collective communication primitive, which disseminates the newly computed document embeddings across all distributed workers.

Furthermore, we developed a custom \textit{Dedup Join} operator, implemented as an efficient CUDA kernel to maximize throughput. This operator scans the user history sequences within the current batch, identifies all occurrences of the recently calculated document IDs, and replaces their stale historical embeddings with the freshly computed counterparts. This ensures that all representations of an item within a single batch are consistent and up-to-date. 
\end{document}